\newcommand{\beq}{  \begin{eqnarray}}
\newcommand{\eeq}{  \end{eqnarray}}

\documentclass[12pt]{iopart}
\usepackage{harvard}  
\usepackage{graphicx}% Include figure files
\usepackage{dcolumn}% Align table columns on decimal point
\usepackage{bm}% bold math

\begin{document}
    	
\title[Cold chemsitry]{Cold ion-atom chemistry driven by spontaneous radiative relaxation: a case study
for the formation of the $ \rm {YbCa^{+}}$ molecular ion}
\author{B. Zygelman$^1$,  Zelimir Lucic$^1$ and Eric R. Hudson$^2$}
\address{$^{1}$ Department of Physics and Astronomy, University of Nevada, Las Vegas, Las Vegas NV 89154, USA}
\address{$^{2}$ Department of Physics and Astronomy, University of California, Los Angeles, California 90095, USA}
\eads{\mailto{bernard@physics.unlv.edu}}
\begin{abstract} {
Using both quantum and semi-classical methods, we calculate the rates for radiative association and charge transfer 
 in cold collisions of ${\rm Yb}^{+}$ with ${\rm Ca}$.  We demonstrate the fidelity of the local optical potential
method in predictions for the total radiative relaxation rates. We find a large variation in the isotope
dependence of the cross sections at ultra-cold gas temperatures. However, at cold temperatures,
 $ \rm  1 mK  < T < 1 K $, the effective spontaneous radiative rates for the different isotopes share a
 common value of about $ 1.5 \times 10^{-15} \, {\rm cm^{3} \, s^{-1}} $. It is
is about five orders of magnitude smaller than the chemical reaction rate measured in [Rellergert et al.,
PRL ${\bm 107}$, 243201 (2011)].}
\end{abstract}
\date{\today}
\maketitle

\section{Introduction}
Advances in the cooling, trapping, and manipulation of ultra-cold
atoms have opened new vistas in our understanding of quantum degenerate matter.
In recent years, laboratory techniques have advanced so that it is now feasible to cool ions and 
explore their interactions with neutral matter in the sub- to milli-Kelvin temperature range. The development of  hybrid, ion-atom
traps~\cite{schmid2010,zipkes2010,zipkes2012,hudson11,hall2011,hall2013} has allowed researchers to explore competing
pathways for reaction between cold atoms and ions, including non-radiative and radiative charge transfer as well as radiative association, in which ions and atoms combine to form a molecule at cold temperatures. Among possible applications for cold ion-atom chemistry are quantum-limited control of chemical reactions and buffer gas cooling of single ion clocks~\cite{zipkes2012}. The aforementioned reactions 
are also important in astrophysical applications~\cite{hudson11,stancil96,zyg98}. Laboratory efforts in measuring accurate rate coefficients of the latter enhances the atomic data base employed in astrophysical models.   

In a recent laboratory study~\cite{hudson11}, Rellergert et al. used a hybrid trap to investigate the interactions between
cold $\rm ^{174}Yb^{+}$ ions with $ \rm ^{40}Ca $ atoms. They observed a large, on the order of $ 2 \times 10^{-10} \, {\rm cm^{3} s^{-1}}$ ,
chemical reaction rate coefficient and, based on a preliminary theoretical estimate, suggested that radiative charge transfer 
was the dominant process behind this rate. However, their theoretical estimate also predicted that nearly half of the chemical reactions should produce YbCa$^{+}$ molecules through radiative association, which disagreed with the experimental observation that the fraction of reactions leading to molecule formation was $\leq0.02$. They suggested this discrepancy could be due to systematic effects, but called for more investigation into the system. The measured rate was several orders of magnitude larger than is typically observed in radiative quenching calculations \cite{Kramers1946,bates51,allison65,cohen82,zyg89,stancil96} for lighter species at higher collision temperatures. To better understand the nature of radiative quenching at these temperatures we \cite{zyg12} estimated the radiative rate employing the local-optical potential method \cite{zyg88} for the total of the two reactions
\beq
&& \rm Yb^{+} + Ca \rightarrow Yb +  Ca^{+} + \hbar \omega  \nonumber \\
&& \rm Yb^{+} + Ca \rightarrow YbCa^{+} + \hbar \omega. 
\label{0.1}
\eeq
Our results are several orders of magnitude smaller than that given in \cite{hudson11} and raises doubts on the suggestion, given in that
paper, concerning the role of radiative relaxation. Our calculations employed the local optical potential method, essentially a semi-classical theory,  whose utility at ultra-low  collision energies has been largely\cite{zhou11} un-tested. In addition, we used molecular data that was gleaned from the illustrations given in \cite{hudson11}. At these energies small details in the potential surfaces and dipole moments can be important. For these reasons, we re-do the calculations here using the original data given in \cite{hudson11}. We compare the predictions of the local-optical potential with the results obtained using the Fermi-Golden-Rule prescription\cite{zyg88}.

The local optical potential method has it roots in the semi-classical theory of radiative association first developed by Kramers and Ter-Haar\cite{Kramers1946}.
In that theory a photon is emitted with energy equal to the energy defect of the two Born-Oppenheimer (BO) potential surfaces
at the internuclear separation in which a transition occurs (see illustration in  figure \ref{fig:fig1}). The efficiency of a transition is determined
by an  Einstein-A coefficient at that internuclear distance. The total rate is then estimated by a, classical, time average over all localized
transitions as the quasi-molecule evolves. This picture was later refined, e.g. see \cite{cohen82}, so that the quantum nature of the entrance channel is fully taken into
account. The relationships between the various semi-classical theories and those obtained using the Fermi-Golden Rule (FGR) methods\cite{sando71} was explored in  \cite{zyg88}. 

In our discussion below we briefly summarize the various theoretical approaches and use them to
calculate the rates for processes (\ref{0.1}). We a provide a rigorous upper bound for the sum of rates given in (\ref{0.1}). Atomic
units are used throughout, unless otherwise indicated. 
\section{Theory}
\subsection{Radiative association}
The cross section for the radiative association process 
\beq
\rm Yb^{+} + Ca \rightarrow YbCa^{+} + \hbar \omega
\label{1.1}
\eeq
where $\hbar \omega $ is the energy of the emitted photon is given by the expression\cite{zyg98}
\beq
&& \sigma_{RA}
 = \sum_{J} \sum_{n} \, \frac{8}{3} \frac{\pi^{2}\omega_{nJ}^{3}}{c^{3} k^{2}} \Bigl [  (J +1)M^{2}_{J+1,J}(k,n) + M^{2}_{J-1,J}(k,n) \Bigr ] \nonumber \\
&& M_{J,J'}(k,n)= \int_{0}^{\infty} dR \, f_{J}(k R) D(R) \phi_{J'n}( R)
\label{1.2}
\eeq
where $D(R)$ is the transition dipole moment between the $ X \,^{2}\Sigma^{+}$ and $ A \, ^{2}\Sigma^{+}$ states of the
$ \rm YbCa^{+}$ molecular ion. $\phi_{nJ}$ is a rho-vibrational eigenstate of the $X \, ^{2}\Sigma^{+}$ ground state, with energy eigenvalue $\epsilon_{nJ}$, 
and is characterized by the angular and vibrational momentum quantum numbers 
$J$, $n$ respectively. $f_{J}(kR)$ is the wavefunction that satisfies the radial Schrodinger equation
\beq
f''_{J}( kR)  - \frac{J(J+1)}{R^{2}} f_{J}( kR) + 2 \mu V_{A} ( R) f_{J}(kR)+k^{2} f_{J}(kR)=0
\label{1.3}
\eeq
where $V_{A}(R)$ is the Born-Oppenheimer (BO) energy of the excited $A ^{2}\Sigma^{+}$ state, $\mu$ is the reduced mass of 
the collision system and $k$ is the wavenumber
for the incident collision partners in that channel. It has the asymptotic form
\beq
f_{J}(k R ) \rightarrow \sqrt{\frac{2 \mu}{\pi k}} \,  \sin(k R -\frac{J \pi}{2}+\delta_{J}), 
\label{1.4}
\eeq
where $\delta_{J}$ is a phaseshift,  as $ R \rightarrow \infty$. The energy of the emitted photon is given by
\beq
\hbar \omega_{nJ} = \frac{\hbar k^{2}}{2\mu}+ V_{A}(\infty) - \epsilon_{nJ} - V_{X}(\infty). 
\label{1.5}
\eeq 
\begin{figure}[ht]
\centering
\includegraphics[width=0.7\linewidth]{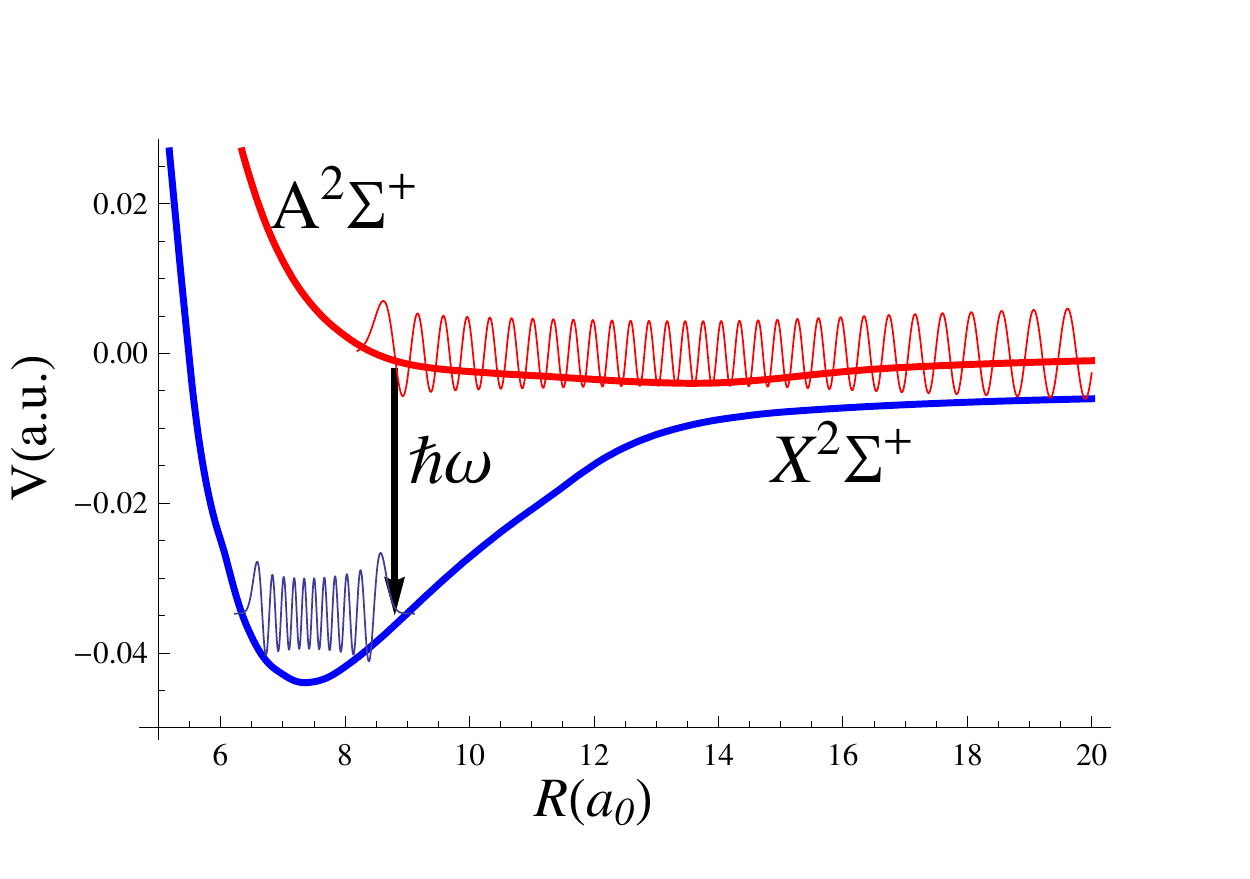}
\caption{\label{fig:fig1}(Color online) Illustration of the BO molecular potential curves (solid thick lines)
 participating in the radiative association
process. In the $A^{2}\Sigma^{+}$ entrance channel the wave function is shown by the light undulating line. The oscillations are
due to the strong polarization force in the entrance channel leading to a potential minimum at $ \rm R \approx 14 a_{0} $.
Association is precipitated by the emission of a photon of energy $\hbar \omega$ near the classical turning point. 
The final bound rho-vibrational state, in the $X^{2}\Sigma^{+}$ channel, is shown by the thin line.}
\end{figure}
\subsection{Radiative charge transfer}
The cross section for the radiative charge transfer process,
\beq
Yb^{+} + Ca \rightarrow Yb+ Ca^{+} + \hbar \omega
\label{1.6}
\eeq
is given by\cite{zyg89}
\beq
&& \sigma_{CT}= \int_{0}^{\omega_{max}} d \omega \, \frac{d \sigma}{d \omega}, \nonumber \\
&& \frac{d \sigma}{d \omega}
 = \sum_{J} \frac{8}{3} \frac{\pi^{2}\omega^{3}}{c^{3} k^{2}} \Bigl [  J M^{2}_{J,J-1}(k,k') + (J+1) M^{2}_{J,J+1}(k,k') \Bigr ]
\label{1.7}
\eeq
where
\beq
&& M_{J,J'}(k,k')= \int_{0}^{\infty} dR \, f_{J}(k R) D(R) f_{J'}(k' R). \nonumber \\
\label{1.8}
\eeq
Here $ f_{J}(k R)$ is a solution to (\ref{1.3}) and $f_{J'}(k' R)$ obeys the  corresponding equation for the ,$X \, ^{2}\Sigma^{+}$,
exit channel with wavenumber and partial wave $k', J'$ respectively. The radial wavefunctions are normalized as in (\ref{1.4})
and
\beq
&& \hbar \omega = \frac{\hbar k^{2}}{2\mu}-\frac{\hbar k'^{2}}{2\mu}+ \Delta E  \nonumber \\
&&  \Delta E \equiv V_{A}(\infty) - V_{X}(\infty).
\label{1.9}
\eeq\
According to (\ref{1.9}) the maximum angular frequency $\omega_{max}$ is given by
\beq
\hbar \omega_{max} = \frac{\hbar k^{2}}{2\mu}+ \Delta E.
\label{1.9a}
\eeq
The sums given by (\ref{1.7}) can be evaluated as in \cite{stancil96}, but here we use a simplified
expression, derived in the Appendix, in which $\sigma_{CT}$ is replaced by its upper bound, i.e.
\beq
\sigma_{CT} < {\tilde \sigma}_{CT} = 
\frac{8}{3} \frac{\pi^{2} \omega^{3}_{max} }{c^{3} k^{2}} \sum_{J} (2J+1) \int_{0}^{\infty}dR f^{2}_{J}(k R) D^{2}(R).
\label{1.10}
\eeq 
\subsection{Optical potential approach}
\begin{figure}[ht]
\centering
\includegraphics[width=0.5\linewidth]{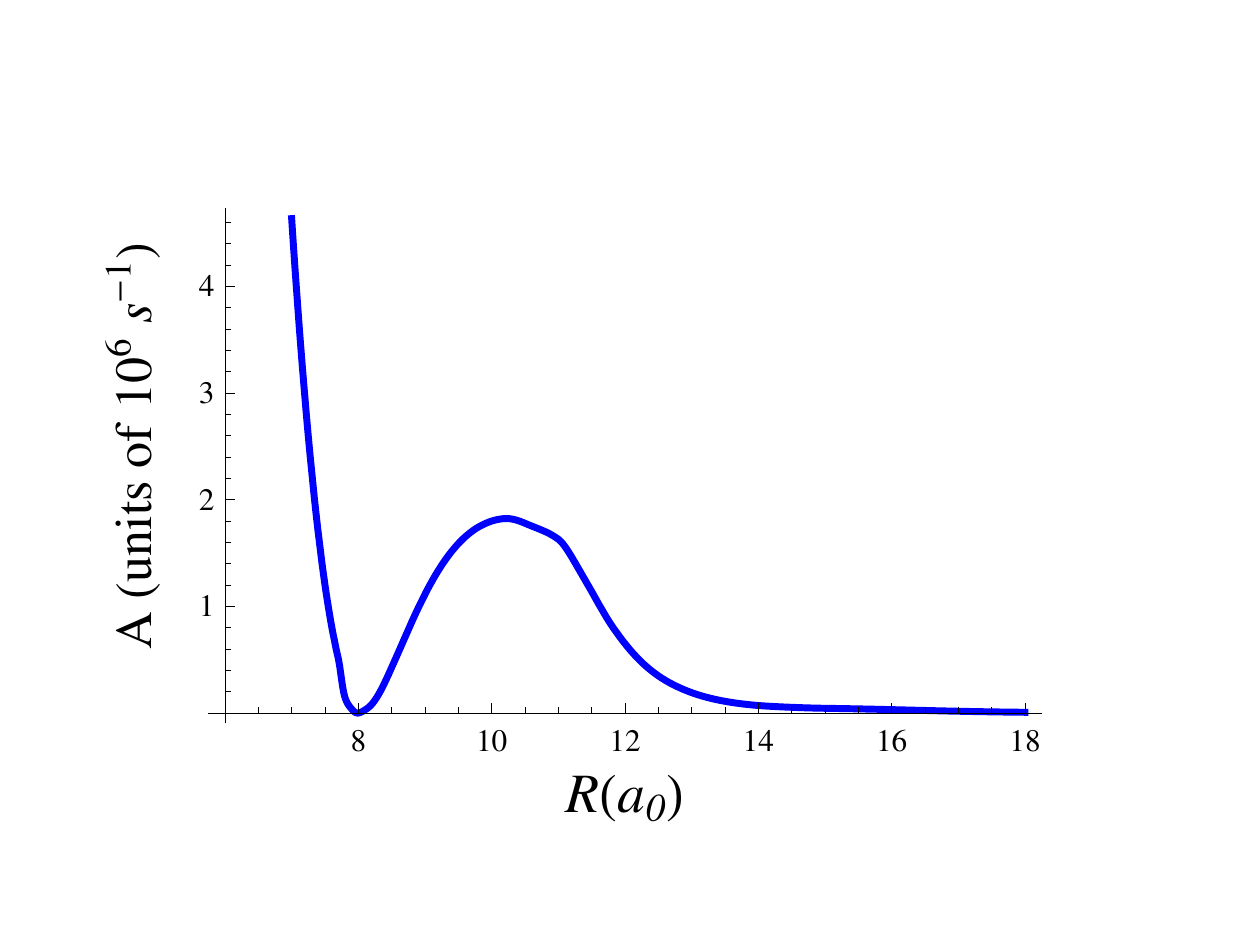}
\caption{\label{fig:fig2}(Color online) Einstein A coefficient as a function of internuclear distance}
\end{figure}

An alternative approach for the calculation of the total radiative loss cross section is given
by the local optical potential method\cite{zyg88}. In it,  the collision system in the incoming $A \, ^{2}\Sigma^{+}$ state experiences,
in addition to the BO energy $V_{A}(R)$, a complex absorptive potential that has the form
\beq
&& V_{opt}= \frac{i A(r)}{2} \nonumber \\
&& A(R) \equiv \frac{4}{3 c^{3}} D^{2}(R) (V_{A}(R)-V_{X}(R))^{3}
\label{1.11}
\eeq
where $A(R)$ is an $R$-dependent Einstein-A coefficient that is illustrated in figure \ref{fig:fig2}.
The cross section for radiative quenching is given by
\beq
\sigma = \frac{\pi}{k^{2}} \sum_{J} (2 J+1) \Bigl ( 1 - \exp(-4 \eta_{J}) \Bigr )
\label{1.12}
\eeq
where $\eta_{J}$ is the imaginary part of the J'th partial wave phase shift $\delta_{J}$ for the radial wave
$f_{J}(k R)$ that satisfies 
\beq
\fl f''_{J}( kR)  - \frac{J(J+1)}{R^{2}} f_{J}( kR) + 2 \mu (V_{A} ( R) + V_{opt}(R)) f_{J}(kR)+k^{2} f_{J}(kR)=0.
\label{1.13}
\eeq

\section{Ultra-cold limit}
In the limit of ultra-cold temperatures in which only the s-wave of the entrance channel participates, the total radiative association
cross section takes the form\cite{zyg01}
\beq
\sigma=\sum_{n} \frac{16 \,\mu \, \pi \, \omega_{n}^{3}}{3 \,c^{3} k} {\Big |} \int_{0}^{R_{0}} dR \, \phi(R) D(R) \phi^{n}_{J=1}(R) 
{\Big |}^{2}
\label{2.1}
\eeq
where $\phi(R) $ is the s-wave solution to (\ref{1.3}) subject to the boundary condition\cite{zyg01}
\beq
d\Phi (R)/dR |_{R_{0}} =1
\nonumber
\eeq
at some, sufficiently large radius $R_{0}$ and
$ \phi^{n}_{J=1}(R) $ are J=1 rho-vibrational states of the $X^{2}\Sigma^{+}$ potential. Because the overlap integral in (\ref{2.1}) is
independent of the incoming wavenumber $k$, (\ref{2.1}) predicts that the association cross section, in the ultra-cold regime, scales
as the inverse of the collision velocity and, therefore, the rate tends to a constant. 

In calculating  $\phi(R)$ one typically matches the numerical solution for $ f_{J}(kR)$  with the asymptotic
form given by expression  (\ref{1.4}). Because of the  polarization potential 
$C_{4}/R^{4}$, one must typically integrate far into the asymptotic region to achieve convergence. Exact solutions for the $C_{4}/R^{4}$ potential are given by
radial Mathieu functions and better convergence can be achieved by employing the latter in the evaluations for the phaseshifts 
e.g \cite{spector64,holzwarth73}. 
\section{Results}
Figure (\ref{fig:fig1}) illustrates the mechanism for radiative association for the $ \rm Yb^{+}$ ion and $\rm Ca$ atom
that approach in the $A^{2} \Sigma^{+}$ electronic BO state. The BO energies where taken from the data of the ab-initio calculations
reported in  \cite{hudson11}. At large internuclear distances this potential has the form
\beq
V_{A}(R) \rightarrow -\frac{C^{A}_{4}}{R^{4}} \quad C^{A}_{4}=78.5. 
\label{3.1}
\eeq
In the incident $A^{2} \Sigma^{+}$ channel the system can relax via the emission of a photon, and in the case of
association, the final state is a bound rho-vibrational level of the $X^{2} \Sigma^{+}$ channel. In radiative charge
transfer the collision partners can exit in that channel, as a re-arranged 
$ \rm Yb$ - $\rm Ca^{+}$ pair.
In the exit channel,
\beq
V_{X}(R) \rightarrow -\frac{C^{X}_{4}}{R^{4}} + \Delta E  \quad C^{X}_{4}=71.5  \quad  \Delta E = -0.0052 
\label{3.1a}
\eeq
as $ R \rightarrow \infty$.
In calculating the radiative association cross sections given by (\ref{1.2}) we need to itemize all bound states supported
by the $X^{2} \Sigma^{+}$ channel. The total number of bound states can be approximated using the JWKB expression
\beq
\fl n=\sum_{L} \, {\rm Floor}\Bigl [ \int_{R_{c}}^{\infty} dR \sqrt{-2 \mu (V_{X}(R)-\Delta E) +\frac{(L+1/2)^{2}}{R^{2}} }-1/2 \Bigr ] \approx 54,803
\label{3.2}
\eeq
where $R_{C}$ is a classical turning point and ${\rm Floor}[x]$ is the integer lower bound of $x$. 
Because of the large reduced mass $\mu$, the number of bound states contributing is much larger than that for association
of lighter species in which typically several hundred rho-vibrational are supported e.g. see \cite{zyg98}.
However, at cold temperatures the centrifugal repulsion in the entrance channel limits the number of partial waves that
participate and so limits, because of the $J \pm 1$ selection rules, the rho-vibrational levels accessed. For example, at a collision
energy corresponding to a temperature of 1 mK, only levels with $ J $ up to the value $\approx 15$ contribute to the association rate.
In figure (\ref{fig:fig2}) we present the results of our calculations for a collision temperature of 1 mK. In that
figure the circles represent the partial wave association cross sections obtained using the FGR expression (\ref{1.7}), the
symbol $X$ in that figure represents the upper limit for total radiative relaxation, which is obtained by adding the association
cross sections (\ref{1.7}) with those given by expression (\ref{1.10}). The square icons represent the cross sections predicted by
expression (\ref{1.12}). It is evident, from this figure, that for $ J < 10$ the optical potential method provides an excellent
approximation for the total cross sections, and for $J>J_{max}$\cite{zyg12},
\beq
J_{max} = \sqrt[4]{8 \,\mu k^{2} \, C^{A}_{4}} = \sqrt[4]{24 \, \mu^{2} \, k_{B} T \, C^{A}_{4}} \approx 12
\label{3.3}
\eeq  
the optical potential method is somewhat less reliable, though still gives reasonable order of magnitude estimates.
$J_{max}$ is the critical angular momentum for which the collision system, approaching in the incident channel at a given energy,
has sufficient collision energy (here given by $3/2 k_{B} T$, where $k_{B}$ is the Boltzmann constant and $T$ is the temperature in Kelvin)
to overcome the centrifugal potential barrier\cite{zyg12}. For larger $J$ tunneling resonances can
access the inner region where the transition dipole moment is non-negligible 
and induce a radiative transition. 
\begin{figure}[ht]
\centering
\includegraphics[width=0.5\linewidth]{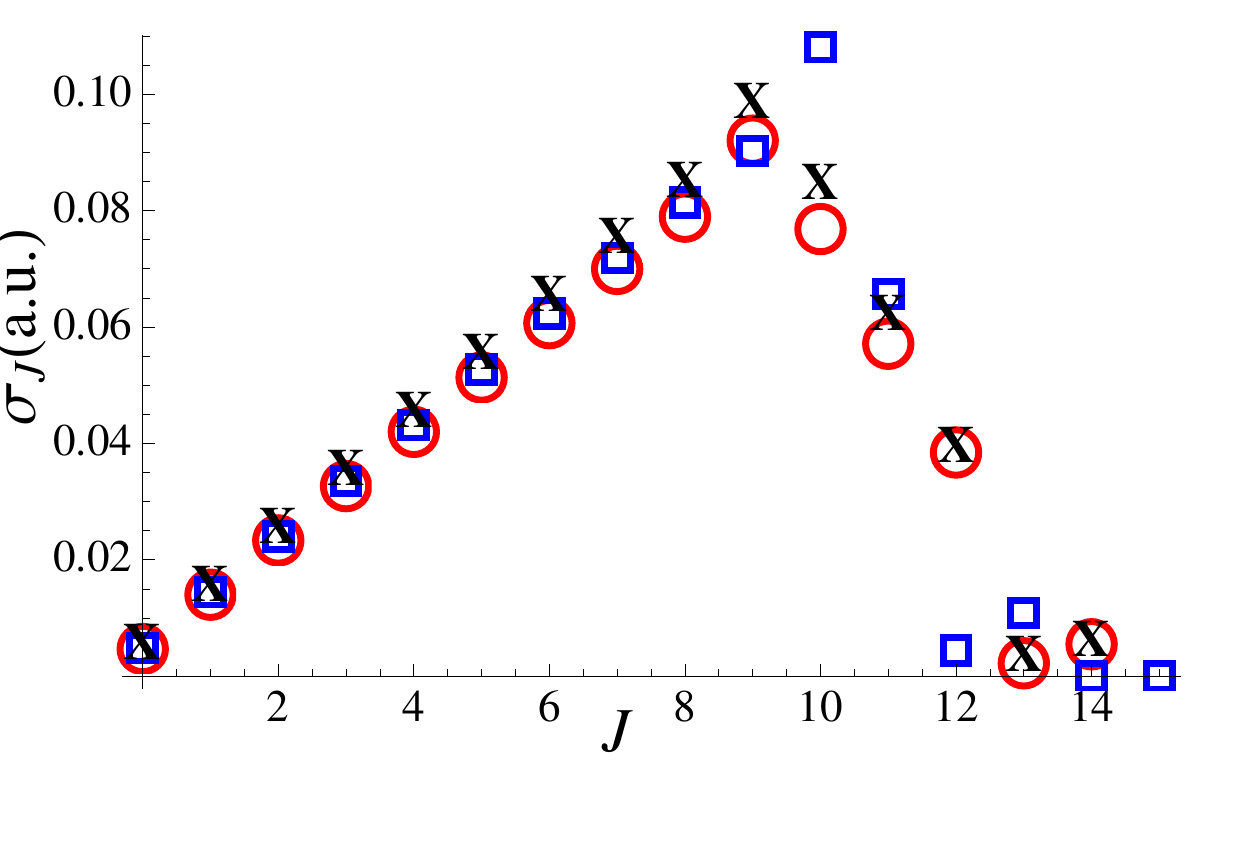}
\caption{\label{fig:fig3}(Color online) Plot of various cross sections as a function of incoming partial wave $J$.
Circles represent data for the radiative association cross sections, $X$'s represent the upper bound for total (association
+ radiative) charge transfer, and squares represent the data for the total radiative relaxation obtained using the optical
potential method.}
\end{figure}
In (\ref{tab:table1}) we tabulate the various cross sections at several representative collision temperatures. In the second
column we itemize the association cross section obtained using the FGR method described above. For 
the radiative charge transfer cross sections, itemized in the third column, we use expression (\ref{1.10}) .
Thus the upper bound for the total radiative relaxation cross sections
are given in column 4. The last column gives the results obtained using the local optical potential method.
\begin{table}
       \caption{\label{tab:table1} Radiative relaxation (RR) cross sections, in units of $a_{0}^{2} $, as a function of
gas temperature.}
\begin{indented}
\item[]
\begin{tabular}{ccccc}
\br
T(K) &  Association & RCT & Total RR &  Optical Potential  \\
\mr
1 pK   & $ 2.88 \times 10^{7} $ &  $2.02 \times 10^{6}$  & $3.08 \times 10^{7} $  &  $3.07 \times 10^{7} $    \\
1 nK   & $ 6.09 \times 10^{4} $  &  4269   &   $6.52 \times 10^{4}$    &  $ 6.48 \times 10^{4} $   \\
$ 1 \rm \mu$   K & 29.23   &  2.049  &   31.28    &   31.11   \\
$ 10 \rm \mu$ K & 6.294  &   0.441  &   6.735   &  6.698  \\
$ 100 \rm \mu$ K & 1.909 &   0.134  &  2.043   &  2.0315   \\
$ 1 $  mK          &  0.626 &   0.0439  &  0.670  &  0.666  \\
$ 10 $  mK         & 0.201 &    0.0132  &  0.214   &  0.214   \\
\br
\end{tabular}
\end{indented}
\end{table}
The table shows that, over the temperature range considered, the local optical potential method predicts cross sections
that are less than the upper bound itemized in column 4. Secondly, the differences between the predictions
of the two theories are small. The optical potential cross section differs
by less than 4 \% from the upper limit values over the entire temperature range, including the ultra-cold region.
We also note that the optical potential method predicts cross sections that are larger than the radiative association
cross sections which underscores an observation cited made in (\cite{zyg89}), that the optical potential
method provides a reliable upper bound for the total (RR) cross section.
In figure (\ref{fig:fig4})  
\begin{figure}[ht]
\centering
\includegraphics[width=0.5\linewidth]{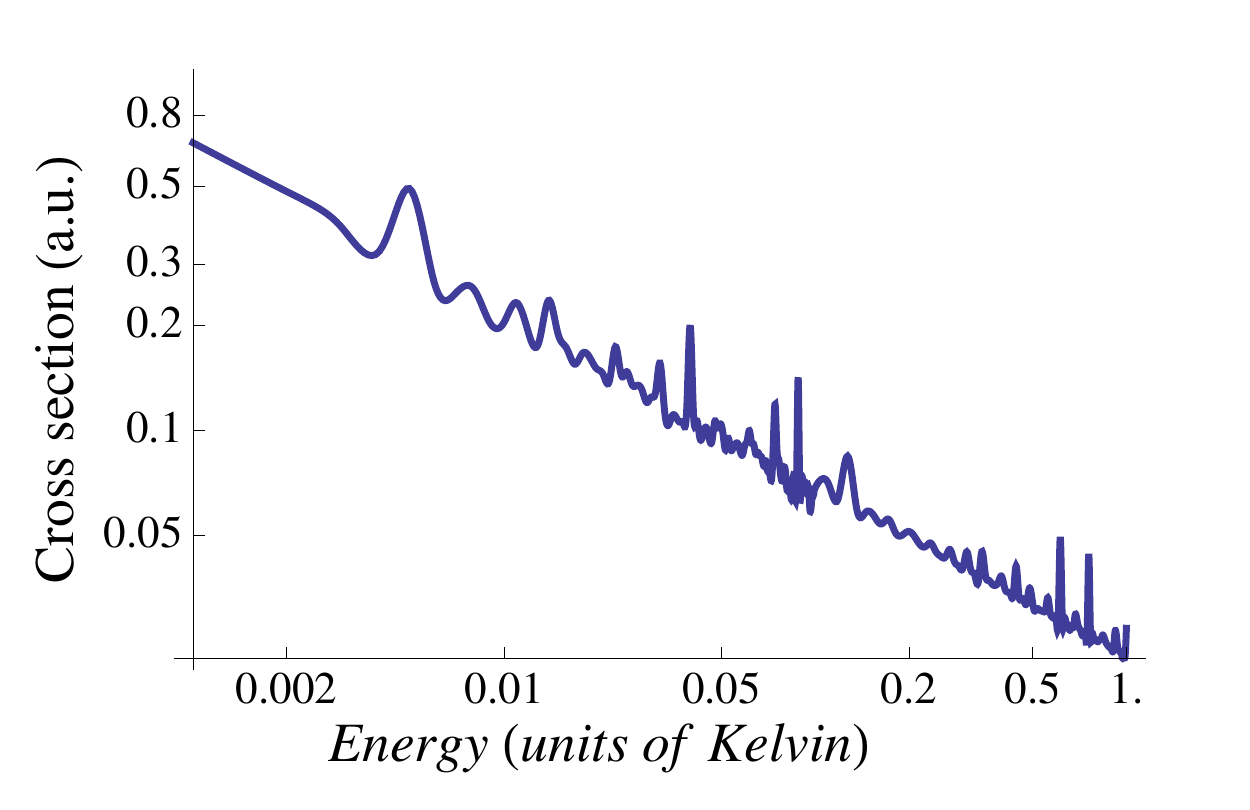}
\caption{\label{fig:fig4}(Color online) Total radiative relaxation cross section as a function of collision energy expressed
as $ \rm {E=\frac{3}{2} k_{B} T }$, where $ {\rm T} $ is the temperature in Kelvin.}
\end{figure}
we plot the total radiative relaxation cross section, obtained using the optical potential method, for the gas temperature
range $ {\rm 1 mK} < {\rm T} < 1 K$. 
Though the optical potential method provides a good approximation for the total radiative relaxation rate, calculation of the photon
emission spectrum requires the use of the FGR method.
In figure (\ref{fig:fig5}) we illustrate the association cross sections $\sigma_{nJ}$, at $\rm T= 1 mK$, for the individual
rho-vibrational levels as function of the frequency of the emitted photon. The structure in the emission pattern, which shows
regions of suppressed and enhanced emission is a result of the oscillations in the incoming wave illustrated in figure (\ref{fig:fig1}).   

In the limit as $ \rm T \rightarrow 0$ we define a complex scattering length for the s-wave solution to
(\ref{1.13}),
\beq
&& a \equiv - \frac{1}{k} \tan \delta(k) =1.64 \times 10^{5}-i \, 1.8591 \nonumber \\
&& {\rm as } \quad k \rightarrow 0.
\label{3.4}
\eeq
Therefore, the total RR cross section, according to the optical potential method, has the limiting value
\beq
\sigma=\frac{4\pi}{k} |Im[a]| = \frac{4\pi}{k} \, 1.859. 
\label{3.5}
\eeq
Defining the rate coefficient 
\beq
k_{RR} \equiv \langle v \sigma \rangle \quad T \rightarrow 0 
\label{3.6}
\eeq
we obtain $ k_{RR} \approx 2.2 \times 10^{-12} \, {\rm cm^{3} \, s^{-1}} $ and is about three orders of magnitude
larger than the corresponding rate in the temperature range $ {\rm 1 mK <  T  < 1 K}$. 
\begin{figure}[ht]
\centering
\includegraphics[width=0.5\linewidth]{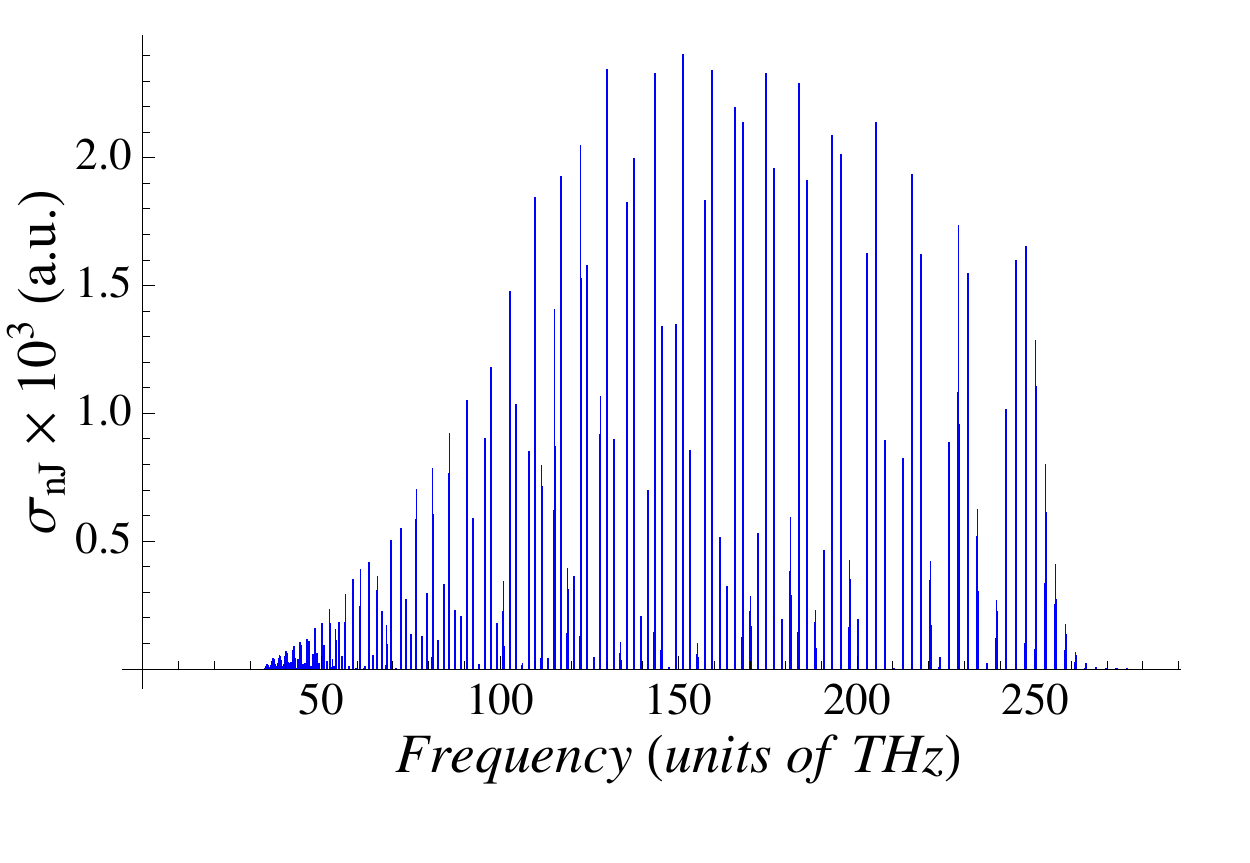}
\caption{\label{fig:fig5}(Color online) Emission spectrum for the radiative association process at a 
gas temperature of $ \rm 1 mK$.}
\end{figure}
In figure (\ref{fig:fig6}) we plot the effective rates $ k \equiv \langle v \rangle \sigma $ for different isotopes
of the $Yb^{+}$ ion. At temperatures $ \rm T > 1 \mu K $  the three rates, corresponding to the isotopes labeled
in that figure, merge to a common value of about $ 1.5 \times 10^{-15} {\rm cm^{3} s^{-1} }$. This feature can
be attributed to Langevin behavior \cite{wannier} which predicts that at low, but high enough that many partial
waves contribute, temperatures ion-atom cross sections scale as the inverse of the collision velocity and therefore the rate 
coefficient tends to a constant. The value of that constant is only weakly dependent on the reduced mass of the collision system 
(e.g. see (2) in \cite{zyg12} ) and that behavior is evident in  figure (\ref{fig:fig6}). In the ultra-cold temperature
regime, where s-wave scattering dominates, the $1/v$ behavior in the cross sections is also operative, e.g. see (\ref{3.4}),
but for a different reason. Whereas in the Langevin regime the cross sections are governed largely by the $C_{4}$ coefficient,
the ultra-cold s-wave phaseshift is also sensitive, as required by Wigner-threshold theory,
to short-range parameters. So the presence of a real, or virtual, bound state near threshold can strongly influence that cross
section. As a consequence, radiative quenching rates which are nearly constant in both the Langevin and ultra-cold regions, 
can suffer rapid variations in the temperature range that adjoins the two territories. This behavior is illustrated in figure (\ref{fig:fig6}) 
by the rate for the $^{174}Yb$ isotope. For this isotope, a bound state near the
threshold leads to significant enhancement in the s-wave cross section at ultra-cold temperatures, the corresponding rate
differs significantly from that in the Langevin region. 
\begin{figure}[ht]
\centering
\includegraphics[width=0.5\linewidth]{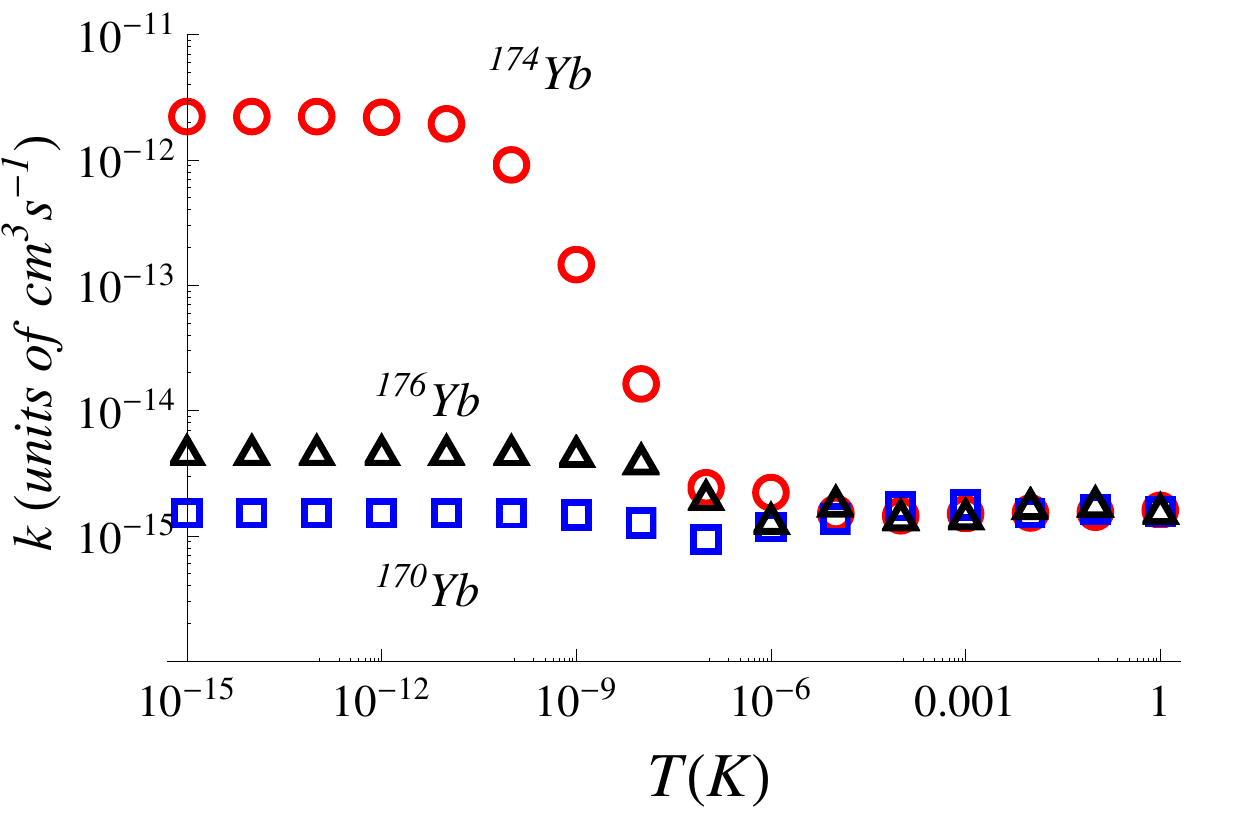}
\caption{\label{fig:fig6} Plot of the effective rate $\rm k \equiv  \langle v \rangle \sigma $, where $\sigma$ is the total radiative
charge transfer cross section, for various isotopes of Ytterbium ions in collisions with cold and ultracold $\rm Ca$ atoms. The horizontal
axis denotes the effective temperature $\rm T$ defined so that the center of mass relative velocity, $\rm v=\sqrt{3 k_{B} T}$ where $\rm k_{B}$ is
the Boltzmann constant.}
\end{figure}
\section{Summary and discussion}
We have presented a computational study of the collision induced radiative processes (\ref{0.1}) at gas temperatures that range
from the cold to ultra-cold regimes. We found that at cold temperatures the total effective rates for theses radiative processes
is no larger than about $ 10^{-15}  \, \rm cm^{3} \, s^{-1}$. We evaluated the fidelity of the local optical potential method\cite{cohen82,zyg88}
in its ability to predict radiative quenching rates, and found that it provides very accurate estimates for the latter even in the
ultra-cold regime. We validated Langevin behavior (which predicts nearly constant rates as a function of temperature) at higher temperatures
but found dramatic departures, and a strong isotope dependence, in the transition from the cold to ultra-cold regimes. At ultra-cold
temperatures the association rate also tends to a constant\cite{zyg01} but for a different reason. Wigner threshold behavior for the s-wave
is strongly influenced by short-range parameters. Thus, there is a collision energy range in which there could be a rapid change in the effective rate,
as illustrated in figure \ref{fig:fig6}. 

This study demonstrates that is is unlikely that the large rate reported in  \cite{hudson11} is solely due to processes (\ref{0.1}). 
%Though the the value of the calculated rate is sensitive to details in the molecular data, we cannot envision a scenario where improvement in the
%quality of the ab-initio data leads to a $10^{5}$ factor enhancement in the rate coefficient. 
In addition to reactions (\ref{0.1}), the system may also undergo the reaction
\beq
 && \rm Yb^{+} + Ca \rightarrow YbCa^{+} (A \, ^{2}\Sigma^{+}) + \hbar \omega. 
\label{4.1}
\eeq
where association proceeds into the weakly bound $(A \, ^{2}\Sigma^{+})$ molecular state in which the collision partners initially approach.
Because process (\ref{4.1}) is driven by a dipole moment that, at large $R$, is proportional to the internuclear distance, one might anticipate a significant rate for it. However, we estimate that this rate is negligible at the temperatures operative in this experiment. Therefore, we conclude that both additional experimental and theoretical studies are necessary in order to understand and reveal cold chemistry of the $\rm Ca^{+} Yb$ system. On the experimental side, revisiting this reaction with new methods recently developed to better probe both the products of the reactions~\cite{Scho12} and the role, if any, of excited electronic states~\cite{Sull12} may help elucidate the relevant pathways. While, on the theoretical front, improved ab initio molecular potentials might help better understand the potential role of non-adiabatic effects in this reaction~\cite{zyg89}. 

\ack
We thank Prof. Svetlana   Kotochigova for use of the ab-initio molecular data.
\appendix{
\section{Bound on RCT cross sections}
According to (\ref{1.7}) - (\ref{1.9}) the frequency of the emitted photon, during an RCT transition, is
\beq
\fl \hbar \omega = \frac{\hbar k^{2}}{2\mu}-\frac{\hbar k'^{2}}{2\mu}+V_{A}(\infty)-V_{X}(\infty) = \frac{\hbar k^{2}}{2\mu}-E'+V_{A}(\infty)-V_{X}(\infty).
\eeq
Now $ d( \hbar \omega ) = - d E' $ 
and $ \hbar \omega_{max} = \frac{\hbar k^{2}}{2\mu}-E'+V_{A}(\infty)-V_{X}(\infty) $, which corresponds to $E'=0$, and
$ \hbar \omega =0 $ for $ E'_{max} =  \frac{\hbar k^{2}}{2\mu} +  V_{A}(\infty)-V_{X}(\infty)$. Therefore (\ref{1.7}) can be 
written as
\beq   
\fl \sigma = \frac{8}{3} \frac{\pi^{2}}{c^{3} k^{2}}    \int_{0}^{E'_{max}} dE' \omega^{3}(E')\,  \Bigl [  J M^{2}_{J,J-1}(k,E') + (J+1) M^{2}_{J,J+1}(k,E') \Bigr ].
\eeq
We have the inequality
\beq
\fl \sigma <  
\frac{8}{3} \frac{\pi^{2} \omega^{3}(E')_{max} }{c^{3} k^{2}} \int_{0}^{\infty} dE' \,
\Bigl [  J M^{2}_{J,J-1}(k,E') + (J+1) M^{2}_{J,J+1}(k,E') \Bigr ].
\eeq
Consider the integral
\beq
\fl && \int_{0}^{\infty} dE'  J M^{2}_{J,J-1}(k,E') = \nonumber \\
\fl &&  \int_{0}^{\infty} dE' \, \int_{0}^{\infty} dR \,
f_{J}(k R) D(R) f_{J-1}(k' R) \, \int_{0}^{\infty} dR' \, f_{J}(k R') D(R') f_{J-1}(k' R').
\eeq
Thus
\beq
\fl \int_{0}^{\infty} dE'  J M^{2}_{J,J-1}(k,E') < \sum_{E'} J M^{2}_{J,J-1}(k,E') = J \int_{0}^{\infty}dR f^{2}_{J}(k R) D^{2}(R).
\eeq
where the sum $ \sum_{E'} $ includes all, bound and continuum states of the exit channel, 
and the second inequality follows from closure properties for the final states
for a given
value of $J$. Therefore we obtain the inequality
\beq
\sigma < \frac{8}{3} \frac{\pi^{2} \omega^{3}(E')_{max} }{c^{3} k^{2}} \sum_{J} (2J+1) \int_{0}^{\infty}dR f^{2}_{J}(k R) D^{2}(R).
\eeq 
}

%\bibliographystyle{jphysicsB}
%\bibliography{bibfile}
\section*{References}
%\begin{thebibliography}{xx}
\begin{harvard}

\harvarditem{{Allison} \harvardand\ {Dalgarno}}{1965}{allison65}
{Allison} D~C~S \harvardand\ {Dalgarno} A  1965 {\em Proceedings of the
  Physical Society} {\bf 85},~845--849.

\harvarditem{Bates}{1951}{bates51}
Bates D~R  1951 {\em Proc. Phys. Soc.} {\bf 64 B},~805.

\harvarditem{{Hall} et~al.}{2011}{hall2011}
{Hall} F~H~J, {Aymar} M, {Bouloufa-Maafa} N, {Dulieu} O \harvardand\
  {Willitsch} S  2011 {\em Physical Review Letters} {\bf 107}(24),~243202.

\harvarditem{{Hall} et~al.}{2013}{hall2013}
{Hall} F~H~J, {Eberle} P, {Hegi} G, {Raoult} M, {Aymar} M, {Dulieu} O
  \harvardand\ {Willitsch} S  2013 {\em ArXiv eprints} .

\harvarditem{{Holzwarth}}{1973}{holzwarth73}
{Holzwarth} N~A~W  1973 {\em Journal of Mathematical Physics} {\bf
  14},~191--204.

\harvarditem{{Kramers} \harvardand\ {Ter Haar}}{1946}{Kramers1946}
{Kramers} H~A \harvardand\ {Ter Haar} D  1946 {\em Bulletin of the Astronomical
  Institutes of the Netherlands} {\bf 10},~137.

\harvarditem{{Ratschbacher} et~al.}{2012}{zipkes2012}
{Ratschbacher} L, {Zipkes} C, {Sias} C \harvardand\ {K{\"o}hl} M  2012 {\em
  Nature Physics} {\bf 8},~649--652.

\harvarditem{Rellergert et~al.}{2011}{hudson11}
Rellergert W~G, Sullivan S~T, Kotochigova S, Petrov A, Chen K, Schowalter S~J
  \harvardand\ Hudson E~R  2011 {\em Phys. Rev. Lett.} {\bf 107},~243201.

\harvarditem{Sando}{1971}{sando71}
Sando K~M  1971 {\em Mol. Phys.} {\bf 21},~439.

\harvarditem{{Schmid} et~al.}{2010}{schmid2010}
{Schmid} S, {H{\"a}rter} A \harvardand\ {Denschlag} J~H  2010 {\em Physical
  Review Letters} {\bf 105}(13),~133202.

\harvarditem{{Schowalter} et~al.}{2012}{Scho12}
{Schowalter} S~J, {Chen} K, {Rellergert} W~G, {Sullivan} S~T \harvardand\
  {Hudson} E~R  2012 {\em Review of Scientific Instruments} {\bf 83},~043103.

\harvarditem{{Spector}}{1964}{spector64}
{Spector} R~M  1964 {\em Journal of Mathematical Physics} {\bf 5},~1185--1189.

\harvarditem{{Stancil} \harvardand\ {Zygelman}}{1996}{stancil96}
{Stancil} P~C \harvardand\ {Zygelman} B  1996 {\em Astrophysical Journal} {\bf
  472},~102.

\harvarditem{{Sullivan} et~al.}{2012}{Sull12}
{Sullivan} S~T, {Rellergert} W~G, {Kotochigova} S \harvardand\ {Hudson} E~R
  2012 {\em Physical Review Letters} {\bf 109},~223002.

\harvarditem{Vogt \harvardand\ Wannier}{1954}{wannier}
Vogt E \harvardand\ Wannier G~H  1954 {\em Physical Review} {\bf 95},~1190.

\harvarditem{{West} et~al.}{1982}{cohen82}
{West} B~W, {Lane} N~F \harvardand\ {Cohen} J~S  1982 {\em Physical Review A}
  {\bf 26},~3164--3169.

\harvarditem{Zhou et~al.}{2011}{zhou11}
Zhou Y, Qu Y~Z, Liu C~H \harvardand\ Liu X~J  2011 {\em Chin. Phys. Lett.} {\bf
  28},~033401--1.

\harvarditem{{Zipkes} et~al.}{2010}{zipkes2010}
{Zipkes} C, {Palzer} S, {Sias} C \harvardand\ {K{\"o}hl} M  2010 {\em Nature}
  {\bf 464},~388--391.

\harvarditem{{Zygelman} \harvardand\ {Dalgarno}}{1988}{zyg88}
{Zygelman} B \harvardand\ {Dalgarno} A  1988 {\em Physical Review A} {\bf
  38},~1877--1884.

\harvarditem{Zygelman et~al.}{1989}{zyg89}
Zygelman B, Dalgarno A, Kimura M \harvardand\ Lane N~F  1989 {\em Phys. Rev. A}
  {\bf 40},~2340--2345.

\harvarditem{{Zygelman} \harvardand\ {Hunt}}{2012}{zyg12}
{Zygelman} B \harvardand\ {Hunt} R  2012 {\em ArXiv e-prints} .

\harvarditem{{Zygelman} et~al.}{2001}{zyg01}
{Zygelman} B, {Saenz} A, {Froelich} P, {Jonsell} S \harvardand\ {Dalgarno} A
  2001 {\em Physical Review A} {\bf 63}(5),~052722.

\harvarditem{{Zygelman} et~al.}{1998}{zyg98}
{Zygelman} B, {Stancil} P~C \harvardand\ {Dalgarno} A  1998 {\em Astrophysical
  Journal} {\bf 508},~151--156.

\end{harvard}

\end{document}